\documentclass[showpacs,showkeys,floatfix,prd,aps]{revtex4}
\usepackage{epsfig} 
\usepackage{longtable} 
\usepackage{bm}
\usepackage{graphicx}
\newcommand{\be}{\begin{equation}}
\newcommand{\dd}{\displaystyle}
\newcommand{\ee}{\end{equation}}
\newcommand{\ba}{\begin{eqnarray}}
\newcommand{\ea}{\end{eqnarray}}

\begin{document}

\topmargin0.5cm 
\begin{titlepage} 
\null 
\vspace{2cm} 
\begin{center} 
\Large\bf   
{\boldmath {$J/\psi$  couplings to charmed resonances and to $\pi$}}
\end{center} 
\vspace{1.5cm} 
 
\begin{center} 
\begin{large}
{A. Deandrea$^{a}$, G. Nardulli$^{b,c}$ and 
A. D. Polosa$^{d}$}\\ 
\end{large}
\vspace{5mm} 
{\it{$^a$Institut de Physique Nucl\'eaire, Universit\'e de Lyon I\\
4 rue E.~Fermi, F-69622 Villeurbanne Cedex, France
\\
$\dd ^b$Dipartimento di Fisica, Universit\`a di Bari, I-70124 Bari, Italia  \\
$^c$I.N.F.N., Sezione di Bari, I-70124 Bari, Italia
\\
$\dd ^d$ CERN - Theory Division, CH-1211 Geneva 23, Switzerland}}
\end{center}

\vskip3truecm
\begin{center} 
\begin{large} 
{\bf Abstract}\\[0.5cm] 
\end{large} 
\parbox{14cm}
{ We present an evaluation of the strong couplings $JD^{(*)}D^{(*)}$ and
$JD^{(*)}D^{(*)}\pi$ by an effective field theory of quarks and mesons.
These couplings are necessary to calculate
$\pi+J/\psi\to D^{(*)}+\bar{D}^{(*)}$ cross sections, 
an important background to
the $J/\psi$ suppression signal in the quark-gluon plasma.
We write down the general
effective Lagrangian and compute the relevant couplings 
in the soft pion limit and beyond.
}  
\end{center}   
\vspace{4cm} 
\noindent 
PACS: 13.25.Gv~12.38.Lg\\ 
\vfil 
\noindent 
BARI-TH 459/03\\ 
CERN-TH/2003-45\\
\vfill 
\eject 
\end{titlepage} 
\newpage 
 
\title{\boldmath $J/\psi$  couplings to charmed resonances and to $\pi$}
\author{A. Deandrea}
\email{deandrea@ipnl.in2p3.fr}
\affiliation{Institut de Physique Nucl\'eaire, Universit\'e Lyon I, 4 rue
E.~Fermi,  F-69622 Villeurbanne Cedex, France} 
\author{G. Nardulli}
\email{giuseppe.nardulli@ba.infn.it}
\affiliation{Dipartimento di Fisica, Universit\`a di Bari and I.N.F.N.,
Sezione di Bari, I-70124 Bari, Italia}
\author{A.D. Polosa}
\email{antonio.polosa@cern.ch}
\affiliation{CERN - Theory Division, CH-1211 Geneva 23, Switzerland}
\preprint{BARI-TH 459/03}
\preprint{CERN-TH/2003-45}
\pacs{13.25.Gv 12.38.Lg}
\keywords{Charmonium interactions; $J/ \psi$ couplings}
\begin{abstract} 
We present an evaluation of the strong couplings $JD^{(*)}D^{(*)}$ and
$JD^{(*)}D^{(*)}\pi$ by an effective field theory of quarks and mesons.
These couplings are necessary to calculate
$\pi+J/\psi\to D^{(*)}+\bar{D}^{(*)}$ cross sections, 
an important background to
the $J/\psi$ suppression signal in the quark-gluon plasma.
We write down the general
effective Lagrangian and compute the relevant couplings 
in the soft pion limit and beyond.
\end{abstract}  
 
\maketitle

\section{Introduction} 

\begin{figure}[ht]
\begin{center}
\epsfig{bbllx=0.5cm,bblly=16cm,bburx=20cm,bbury=23cm,height=4.truecm,
        figure=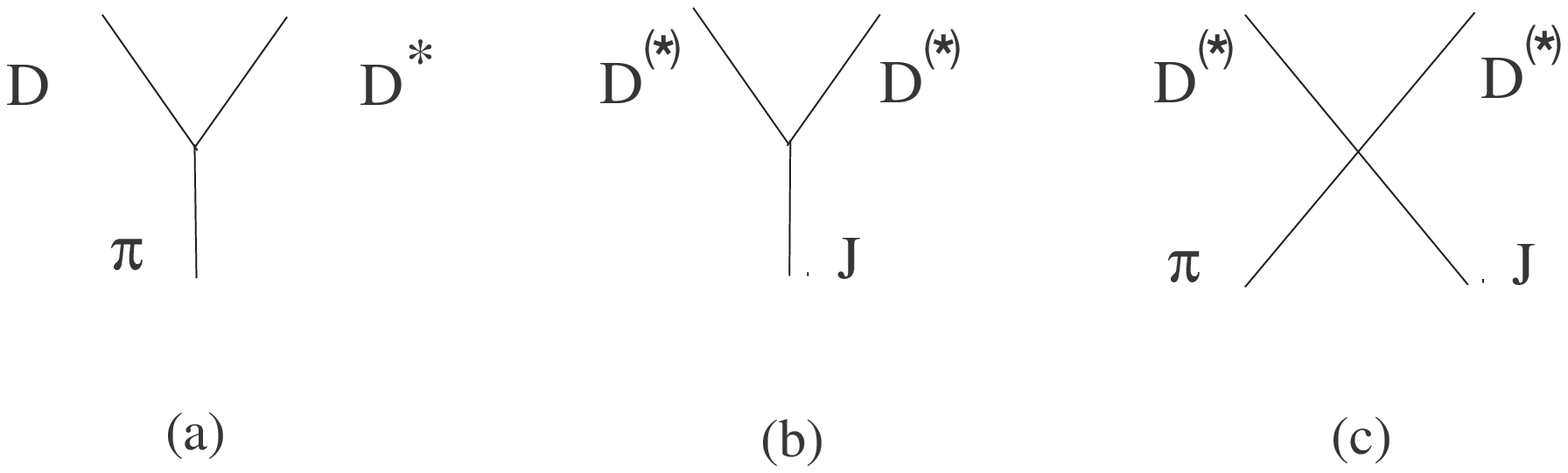}\vskip2cm
\caption{\label{fig.0} \footnotesize  Couplings involved in the
tree level calculations of the processes (\ref{1}).}
\end{center}
\end{figure}

This paper is devoted to the study
of the strong couplings of $J/\psi$, low mass charmed mesons and pions. The
interest of this study stems from the possibility that $J/ \psi$ absorption
processes of the following type 
\be 
\pi +J/\psi\to D^{(*)}+\bar{D}^{(*)}
\label{1} 
\ee 
play an important role in the relativistic
heavy ion scattering. Since a decrease of the $J/\psi$ production  
might signal the formation of Quark Gluon Plasma (QGP) in a
heavy ion collision, it is useful
to have reliable estimates of the cross sections for the processes (\ref{1})
that provide an alternative way to reduce the $J/\psi$ production rate.
Previous studies of these effects can be found in
\cite{cina1,c0,Deandrea:2002kj}. The relevant couplings needed
to compute (\ref{1}) are depicted in Fig.~\ref{fig.0}.

Besides the ${DD^*\pi}$ coupling, see Fig.~\ref{fig.0}a, whose
coupling constant $g_{D^*D\pi}$ has been  theoretically estimated
\cite{theor}, \cite{report} and experimentally investigated
\cite{exp}, to compute the amplitudes (\ref{1}) one would need
also the $JD^{(*)}D^{(*)}$, see Fig.~\ref{fig.0}b, and the
$J\,D^{(*)}D^{(*)}\pi$ couplings, see Fig.~\ref{fig.0}c. In an
effective Lagrangian approach the latter couplings provide direct
four-body interactions, while the former enter the amplitude {\it
via} tree diagrams with the exchange of a charmed particle $D^{(*)}$
in the $t-$channel. These couplings have been estimated by
different methods, that are, in our opinion, unsatisfactory. For
example the use of the $SU_4$ symmetry puts on the same footing
the heavy quark $c$ and the light quarks, which is at odds with
the results obtained within the Heavy Quark Effective Theory
(HQET), where the opposite approximation $m_c\gg \Lambda_{QCD}$ is
used (for a short review of HQET see \cite{report}). Similarly,
the rather common approach based on the Vector Meson Dominance
(VMD) should be considered critically, given the large
extrapolation $p^2=0\to m_{J/\psi}^2$ that is involved. A
different evaluation, based on QCD Sum Rules can be found in
\cite{qcdsr} and presents the typical theoretical uncertainties of
this method. In this note we will use a different approach, based on
the Constituent Quark Meson (CQM) model \cite{nc},\cite{nc1},
which takes into account explicitly the
HQET symmetries. We gave a preliminary report of the present study
in \cite{Deandrea:2002kj} where we presented the results for some
of the couplings (\ref{1}). Here we complete the analysis and
compare our results with the existing literature.

\section{Constituent Quark Meson model\label{sec2}}
The CQM model is a quark-meson model arising from an extension of 
Ref. \cite{Ebert:1994tv} to exploit fully HQET and chiral symmetries in 
the interactions of heavy and light mesons. A survey of these methods can 
be found in \cite{nc},\cite{nc1}. Here we
limit to those aspects of the model that are relevant for the interactions of
$J/\psi$, low mass charmed mesons and pions. The model is an effective field
theory whose effective fields are light and heavy quarks as well as light and
heavy mesons.  The Feynman rules for the model are explicitly written down in
\cite{nc}, \cite{nc1}. The transition amplitudes containing light/heavy mesons
in the initial and final states as well as the couplings of the heavy mesons to
hadronic currents are computable via ù quark loop diagrams where mesons enter
as external legs. The model is relativistic and incorporates, besides the heavy
quark symmetries, also the chiral symmetry of the light quark sector.

To show an example of the calculation in the CQM model we consider the 
Isgur-Wise (IW) function defined e.g. by
\be
\langle D(p^\prime)|\bar c\gamma^\mu c|D(p) \rangle = 
m_D\xi(\omega)(v+v^\prime)^\mu
\ee 
where $p^\mu=m_D v^\mu$, $p^{\prime\,\mu}=m_D v^{\prime\,\mu}$ and
$\omega=v\cdot v^\prime$.
We note that  the Isgur-Wise function obeys the
normalization condition $\xi(1)=1$, arising from  the flavor symmetry of the
HQET.  The explicit definition of the Isgur-Wise form factor is:
\be 
\langle H(v^\prime)|\bar c \gamma_\mu c |H(v)\rangle=
-\xi(\omega){\rm Tr}\left(\bar H\gamma_\mu H\right)\ . 
\ee 
Here $H$ is the multiplet containing
both the $D$ and the $D^*$ mesons   \cite{report}: \be H=\frac{1+\gamma\cdot
v}{2}(-P_5\gamma_5+ \gamma\cdot P)\ ,\ee and $P_5,\,P^\mu$ are annihilation
operators for the charmed mesons. One gets \be \xi(\omega)(v+v^\prime)^\mu=
Z_H\frac{3i}{16\pi^4}\int d^4\ell \frac{{\rm Tr}\left[(\gamma\cdot\ell+m)
\gamma_5 (1+\gamma\cdot v^\prime)\gamma_\mu (1+\gamma\cdot v)\gamma_5\right]}{4
(\ell^2-m^2)(v\cdot \ell+\Delta_H)(v^\prime\cdot\ell-\Delta_H)}\ , \ee where $
v$ and $v^\prime$ are the $4-$velocities of the two heavy quarks that are
equal, in the infinite quark mass limit, to the hadron velocities,\be
\frac{1+\gamma\cdot v}{2(v\cdot \ell+\Delta_H)} \ee is the heavy quark
propagator of  the HQET,  and $\Delta_H=m_D-m_c=v\cdot k$ in  the limit
$m_c\to\infty$; $k$ is the meson residual momentum, defined by $p^\mu=m_c v^\mu
+ k^\mu$. The numerical value of $\Delta_H$ is  in the range $0.3-0.5$~GeV
\cite{nc}. If we consider a $D^*$ meson instead of a $D$, a factor $-\gamma_5$
must be substituted by $\gamma\cdot\epsilon$, $\epsilon$ being the polarization
of $D^*$. The constant $Z_H$ is the heavy meson field wavefunction
renormalization constant giving the strength of the quark-meson coupling (more
precisely the coupling  is $\sqrt{Z_H m_{D}}$). $Z_H$ is computed and tabulated
in~\cite{nc}. One gets for the IW function: 
\be 
\xi(\omega)=Z_H\left[
\frac{2}{1+\omega} I_3(\Delta_H)+
\left(m+\frac{2\Delta_H}{1+\omega}I_5(\Delta_H,\Delta_H,\omega)\right)
\right]\ , 
\ee 
where the $I_i$ integrals are given by: 
\ba 
I_3(\Delta) &=& -\frac{iN_c}{16\pi^4} \int^{\mathrm {reg}} 
\frac{d^4k}{(k^2-m^2)(v\cdot k +
\Delta + i\epsilon)}\nonumber \\ &=&{N_c \over {16\,{{\pi }^{{3/2}}}}}
\int_{1/{{\Lambda}^2}}^{1/{{\mu }^2}} {ds \over {s^{3/2}}} \; e^{- s( {m^2} -
{{\Delta }^2} ) }\; \left( 1 + {\mathrm {erf}}
(\Delta\sqrt{s}) \right)\\
I_5(\Delta_1,\Delta_2,\omega) &= & \frac{iN_c}{16\pi^4} \int^{\mathrm {reg}}
\frac{d^4k}{(k^2-m^2)(v\cdot k + \Delta_1 +
i\epsilon ) (v'\cdot k + \Delta_2 + i\epsilon )} \nonumber \\
 & = & \int_{0}^{1} dx \frac{1}{1+2x^2 (1-\omega)+2x
(\omega-1)}\times\nonumber\\ &&\Big[
\frac{6}{16\pi^{3/2}}\int_{1/\Lambda^2}^{1/\mu^2} ds~\sigma \;
e^{-s(m^2-\sigma^2)} \; s^{-1/2}\; (1+ {\mathrm {erf}}
(\sigma\sqrt{s})) +\nonumber\\
&&\frac{6}{16\pi^2}\int_{1/\Lambda^2}^{1/\mu^2} ds \; e^{-s\sigma^2}\;
s^{-1}\Big], 
\ea 
where
\begin{equation}
\sigma(x,\Delta_1,\Delta_2,\omega)={{{\Delta_1}\,\left( 1 - x \right)  +
{\Delta_2}\,x}\over {{\sqrt{1 + 2\,\left(\omega -1 \right) \,x +
2\,\left(1-\omega\right) \,{x^2}}}}}.
\end{equation}
The ultraviolet cutoff $\Lambda$, the
infrared cutoff $\mu$ and the light constituent mass $m$ are fixed in the
model~\cite{nc} to be $\Lambda=1.25$~GeV, $\mu=0.3$~GeV and $m=0.3$~GeV. 
Other integrals to be used later are 
\ba 
I_2 &=& -\frac{iN_c}{16\pi^4}\int^{\mathrm
{reg}} \frac{d^4k}{(k^2-m^2)^2}= {N_c \over {16\,{{\pi
}^{{2}}}}}\Gamma\left(0,\frac{m^2}{\Lambda^2},
\frac{m^2}{\mu^2}\right)\\
I_4(\Delta)&=&\frac{iN_c}{16\pi^4}\int^{\mathrm {reg}} \frac{d^4k}{(k^2-m^2)^2
(v\cdot k + \Delta + i\epsilon)} \nonumber\\ &=&\frac{N_c}{16\pi^{3/2}}
\int_{1/\Lambda^2}^{1/\mu^2} \frac{ds}{s^{1/2}} \; e^{-s(m^2-\Delta^2)} \;
[1+{\mathrm {erf}}(\Delta\sqrt{s})] 
\ea
The calculation of the $g_{D^*D\pi}$ coupling constant for the matrix element 
\be
\langle\pi^+(q)D^0(p)|D^{*+}(p',\epsilon)\rangle=ig_{D^*D\pi}
\epsilon\cdot q
\ee
in the CQM proceeds along similar lines and can be found in \cite{nc}. We
reproduce for reference's sake since it is an important element of the
amplitudes (\ref{1}). In the soft pion limit (spl) one gets \cite{nc}: 
\be
g_{D^*D\pi}^{spl}=13\pm 1~. 
\label{ddp} 
\ee 
The experimental situation is as follows: CLEO results~\cite{exp} give
$g=0.59\pm 0.07\pm 0.01$ where $g$ is related to the constant in eq. 
(\ref{ddp}) by $g_{D^*D\pi}^{spl}=2m_D g/f_\pi$ ($f_\pi\approx 130$ MeV); on 
the other hand, in general, QCD sum rules predict smaller values, see for a 
review \cite{report}.

\section {$JD^{(*)}D^{(*)}$ couplings}
\begin{figure}[ht]
\begin{center}
\epsfig{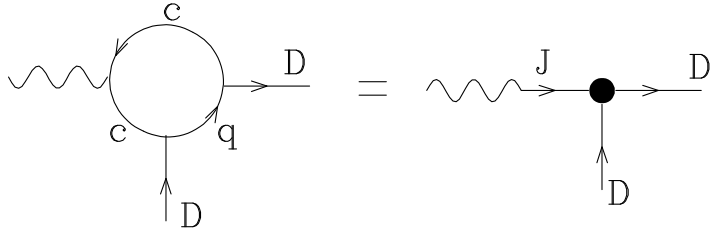}
\caption{\label{fig.1} \footnotesize The Vector Meson Dominance equation giving
the coupling of $J/\psi$ with $D,D^*$ in terms of the Isgur-Wise function
$\xi$. The function $\xi$ on the l.h.s. is computed by a diagram with a quark
loop. The coupling of each $D^{(*)}$ meson to quarks is given by $\sqrt{Z_H
m_D}$.}
\end{center}
\end{figure}

The calculation of the Isgur-Wise
we have described above is a crucial ingredient to the computation
of the $JD^{(*)}D^{(*)}$ vertexes of Fig.~\ref{fig.0}b. It
corresponds to the evaluation of the l.h.s. of Fig.~\ref{fig.1},
while,  {\it via} a VMD ansatz, the r.h.s. gives the desired
coupling. Concerning the use of the VMD for the charm system 
one has to note that it is not based on the 
hypothesis that all higher order resonances give contributions smaller 
than the J/$\psi$, but on the fact that the higher states 
give contributions of alternating sign that tend to cancel. 
This sign difference follows from an evaluation via the saddle point 
method of the WKB approximation \cite{altern}.

The Isgur-Wise function can be computed for any value of $\omega$ and not only
in the region $\omega>1$, which is experimentally accessible $via$ the
semileptonic $B\to D^{(*)}$ decays; $\omega$ is related to the meson momenta by
\be \omega=\frac{p_1^2+p_2^2-p^2}{2 \sqrt{p_1^2p_2^2}}\,,\label{14}\ee where
$p_1,\,p_2=$ momenta of the two $D$ resonances.

Let us now consider the r.h.s. of the equation depicted in Fig.~\ref{fig.1}.
For the coupling of $J/\psi$ to the current we use the matrix element \be
\langle 0|\bar c \gamma^\mu c |J(q,\eta)\rangle=f_Jm_{J}\epsilon^\mu \ee with
$f_J=0.405\pm 0.014$ GeV. As to the strong couplings $JD^{(*)}D^{(*)}$, the
model in Fig.~\ref{fig.1} gives the following effective Lagrangians 
\ba 
{\cal L}_{JDD}&=&ig_{JDD}\left(\bar{D}
{\stackrel{\leftrightarrow}{\partial}}_{\nu}D\right)J^\nu\ , \\ {\cal
L}_{JDD^*}&=&ig_{JDD^*}\epsilon^{\mu\nu\alpha\beta}J_{\mu}
\partial_{\nu}\bar D\partial_{\beta}D^*_\alpha\ ,\\
&&\cr
{\cal L}_{JD^*D^*}&=& ig_{JD^*D^*}\Big[ \bar{D}^{*\mu}\left(
{\partial}_{\mu}D^*_\nu\right)J^\nu - {D}^{*\mu} \left( {\partial}_{\mu}\bar
D^*_\nu\right)J^\nu \cr &-& \left(
 \bar{D}^{*\mu}{\stackrel{\leftrightarrow}{\partial}}_{\nu}
D^*_\mu\right)J^\nu \Big]\ . 
\ea 
Here $D^{(*)}\bar D^{(*)}$ can be any of the
pairs $D^{0(*)}\bar D^{0(*)}$, $D^{+(*)} D^{-(*)}$ or $D_s^{(*)} \bar
D_s^{(*)}$ (neglecting $SU_3$ breaking effects). As a consequence of the spin
symmetry of the HQET we find: 
\be 
g_{JD^*D^*}=g_{JDD}\,,~~~~~~~~~~
g_{JDD^*}=\frac{g_{JDD}}{m_D}\ ,
\ee 
while the VMD ansatz gives:
\be 
\label{eq:coupling1}
g_{JDD}(p_1^2,\,p_2^2,\,p^2)=\frac{m^2_{J}-p^2}{f_J
m_{J}}\xi(\omega) \ .\ee Since $g_{JDD}$ has no zeros, eq.
(\ref{eq:coupling1}) shows that $\xi$ has a pole at $p^2=m^2_{J}$,
which is what one expects on the basis of dispersion relations
arguments. The CQM evaluation of $\xi$ does show a strong peak for
$p^2\approx(2m_c)^2$, even though, due to $\displaystyle {\cal
O}\,\left({1}/{m_c}\right)$ effects, the location of the
singularity is not exactly at $p^2=m^2_{J}$.

\begin{figure}[ht!]
\begin{center}
\epsfig{height=4truecm,figure=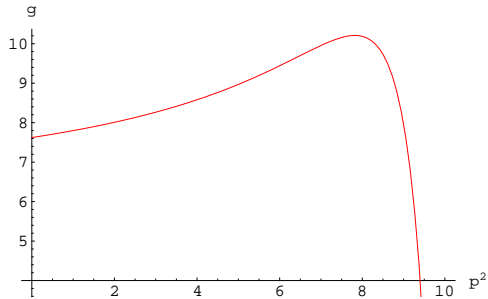} \caption{\label{fig.2} \footnotesize
The $p^2$ dependence of $g=g_{JDD}(m_D^2,\,m^2_D,\,p^2)$, showing the almost
complete cancellation between the pole of the Isgur-Wise function and the
kinematical zero.  Units are  GeV$^2$ for $p^2$.}\end{center}
\end{figure}

This is shown in Fig.~\ref{fig.2} where we plot $g_{JDD}(p_1^2,\,p_2^2,\,p^2)$
for on shell $D$ mesons, as a function of $p^2$ (the plot is obtained for
$\Delta_H=0.4$~GeV and $Z_H=2.36~$GeV$^{-1}$). For $p^2$ in the range $(0,4)$
GeV$^2$, $g_{JDD}$ is almost flat, with a value \be g_{JDD}=8.0\pm 0.5\
.\label{gjdd}\ee For larger values of $p^2$ the method is unreliable due to the
above-mentioned incomplete cancellation between the kinematical zero and the
pole. Therefore, we extrapolate the smooth behavior of $g_{JDD}$ in the small
$p^2$ region up to $p^2=m^2_{J}$ and assume the validity of the result
(\ref{gjdd}) also for on-shell $J/\psi$ mesons. On the other hand in the
$p_1^2,p^2_2$ variables we find a behavior compatible with that produced by a
smooth form factor. Also the result from Ref. \cite{cina1} in the Table, is 
based on a VMD assumption; a previous determination based on the same 
assumption is \cite{matinyan}, with similar results ($g_{JDD}=7.7$).

In Table 1 we compare our results with those
of other authors. We observe that our  results for $g_{JDD}$ and $g_{JDD^*}$
agree with the outcomes of Ref.~\cite{cina2} and with the QCD sum rule analysis
of \cite{qcdsr}; in particular the smooth behavior of the form factor found in
\cite{qcdsr} for $g_{JDD}$ agrees with our result. This is not surprising, as
\cite{cina2} uses a VMD model as well. 

As for the  QCD sum rules calculation,
it involves a perturbative part and a non perturbative contribution, which is
however suppressed. The perturbative term has its counterpart in CQM in the
loop calculation of Fig.~\ref{fig.1} and the overall normalization should agree
as a consequence of the Luke's theorem. On the other hand we differ from 
Ref.~\cite{cina1} for a sign and from \cite{haglin} both in sign 
and in magnitude. The sign difference may be due to an overall phase in
the $J/\Psi$ wavefunction. It is however of no effect in the computation
of the  $J/\Psi$ absorption cross section, which is
the main application of the present calculation.
Finally  Ref.~\cite{cina2} obtains a value for $g_{JDD^*}$ using results from
the decay $J/\psi\to\rho\pi$. This seems to us too strong an assumption due to
the fact that  $J/\psi\to\rho\pi$ could proceed {\it via} gluonic decay of the
$J/\psi$, which is not the case for $J/\psi\to D D^*$.
\begin{center}
\begin{tabular}{|c|c|c|c|c|c|}
\hline   Coupling &
Our work& Ref.~\cite{cina1}& Ref.~\cite{haglin}&Ref.~\cite{cina2}&
Ref.~\cite{qcdsr}\\
\hline
 $g_{JDD}$&$ 8.0\pm 0.5$ &-7.64 &$-4.93$ &7.71&7.36\\
 $g_{JDD^*}$(GeV$^{-1}$)&$ 4.05\pm 0.25$& -& &$8.02\pm 0.62$&-\cr
 $g_{JD^*D^*}$&$ 8.0\pm 0.5$&-7.64 & $-4.93$&7.71&-\cr
 \hline
\end{tabular}
\end{center}
{\footnotesize Table 1. Comparison of theoretical results for the couplings
$g_{JDD}$, $g_{JDD^*}$ and  $g_{JD^*D^*}$. Ref.~\cite{cina1} and 
Ref.~\cite{cina2} use a VMD model similar to the one used in the 
present paper for
the couplings $g_{JDD}$, $g_{JD^*D^*}$. For $g_{JDD^*}$ Ref.~\cite{cina2} uses
VMD together with data from relativistic quark model \cite{Colangelo:1994jc} to
get the coupling of a hadronic current to $D$ and $D^*$. Ref.~\cite{haglin}
uses a chiral model to compute the coupling constants $g_{JDD}$ and
$g_{JD^*D^*}$. The coupling $g_{JDD^*}$ is not included in  Ref.~\cite{cina1}.
Ref.~\cite{qcdsr} is based on QCD sum rules; the result we report for $g_{JDD}$
is computed at the same value $p^2=2$ GeV$^2$ as in our work.} \vskip.3cm

\section{$JD^{(*)}D^{(*)}\pi$ couplings} Let us now consider the
$JD^{(*)}D^{(*)}\pi$ couplings of Fig.~\ref{fig.0}c. We write the
effective Lagrangians for the $J\pi^+D^-D^0$ coupling (other
couplings can be obtained by use of symmetries): 
\ba 
{\cal L}_{JDD\pi}&=&\frac{g_0}{m_D}
\epsilon^{\mu\nu\alpha\beta}J_{\mu}(\partial_{\nu}\pi)
\partial_{\alpha}D\partial_{\beta}\bar{D}\,,\\ 
{\cal L}_{J{\bar D}^* D\pi}&=& 
- \,J_{\mu}(\partial_\nu\pi)
\Big\{g_1\left[g^{\mu\nu}{\bar D}^*_\lambda\partial^\lambda
\,D+2\,D\,\partial^\mu{\bar D}^{*\nu}  \right]
 -g_2 \left[{\bar D}^{*\mu}\partial^\nu+ D\partial^\nu{\bar
D}^{*\mu}\right] \cr
 &+&  \frac{g_3}{m_{D}^2}
\left[\left(\partial^\nu{\bar D}^{*\lambda} \right)
\partial^\mu\partial_\lambda D-
\left(\partial^\nu\partial_\lambda D\right)
\partial^\mu {\bar D}^{*\lambda}
\right] \Big\}~+~h.c.\,,\\{\cal L}_{J{\bar D}^* D^*\pi}&=& m_D
J^{\mu}(\partial^\nu\pi) \Big\{ g_4\,\epsilon_{\mu\nu\rho\lambda}
{D}^{*\rho}{\bar D}^{*\lambda}+
\frac{g_5}{m_D^2}\epsilon_{\mu\alpha\beta\lambda}
\left(\partial^\alpha\partial_\nu{D}^{*\beta}\right) {\bar D}^{*\lambda}
\cr&-&\frac{ g_6}{m^2_D}\epsilon_{\mu\alpha\beta\lambda}
\left(\partial_\nu{D}^{*\alpha}\right)
\partial^\lambda{\bar D}^{*\beta}
\cr &+&\frac{ g_7}{m^2_D} \Big[\epsilon_{\mu\alpha\nu\beta}
\left(\partial_\lambda\partial^\alpha {D}^{*\beta}\right) 
{\bar D}^{*\lambda} -2\,\epsilon_{\mu\alpha\lambda\beta} 
\left(\partial^\alpha {D}^{*\lambda}\right)\partial^\beta
{\bar D}^{*\nu} \cr &+&2\,\epsilon_{\mu\nu\alpha\beta}{D}^{*\lambda}
\partial_\lambda\partial^\alpha
{\bar D}^{*\beta}
+\epsilon_{\alpha\nu\beta\lambda}\left(\partial^\alpha{D}^*_{\mu}\right)
\partial^\lambda
{\bar D}^{*\beta} \cr&+& g_{\mu\nu}
\epsilon_{\alpha\beta\rho\lambda}\left(\partial^\alpha{D}^{*\beta}\right)
\partial^\lambda
{\bar D}^{*\rho}-2\epsilon_{\nu\alpha\beta\lambda}{D}^{*\alpha}
\partial^\lambda\partial_\mu{\bar D}^{*\beta} \Big]
\cr&+& \frac{g_7^\prime}{m^2_D}\,\epsilon_{\mu\alpha\beta\lambda}\left(
\partial^\alpha{D}^{*}_\nu\right)\partial^\lambda{\bar D}^{*\beta}\cr
&+&\frac{g_8}{m^2_D}\,\epsilon_{\mu\alpha\beta\lambda}
{D}^{*\alpha}\partial^\lambda\partial_\nu{\bar D}^{*\beta}+
\frac{g_9}{m^2_D}\,\epsilon_{\mu\alpha\beta\lambda}\left(
\partial^\beta{D}^{*\alpha}\right)\partial_\nu{\bar D}^{*\lambda}\cr
&+& \frac{g_1}{m^2_D}\,\epsilon_{\nu\alpha\beta\lambda}\left(
\partial^\alpha\partial_\mu{D}^{*\beta}\right){\bar D}^{*\lambda}+
\frac{g_7^{\prime\prime}}{m^2_D}\,\epsilon_{\nu\alpha\beta\lambda}\left(
\partial^\alpha{D}^{*\beta}\right)\partial^\lambda{\bar D}^{*}_\mu\cr
&-& \frac{g_9}{m^4_D}\,\left[\epsilon_{\mu\alpha\beta\lambda}\left(
\partial^\alpha\partial_\nu{D}^{*\rho}\right)
\partial^\lambda\partial_\rho{\bar D}^{*\beta}+
\epsilon_{\alpha\rho\beta\lambda}\left(
\partial^\alpha\partial_\nu{D}^{*\rho}\right)
\partial^\lambda\partial_\mu{\bar D}^{*\beta}
\right]\cr&+&\frac{g_9^\prime}{m^4_D}
\left[\epsilon_{\mu\alpha\beta\lambda}\left(
\partial^\alpha{D}^{*\rho}\right)
\partial_\rho\partial_\nu\partial^\lambda{\bar D}^{*\beta}+
\epsilon_{\alpha\rho\beta\lambda}\left(
\partial^\alpha{D}^{*\rho}\right)
\partial^\lambda\partial_\mu\partial_\nu{\bar D}^{*\beta}
\right]\Big\}.
\ea 
While in these formulae 13 coupling constants appear, the
number of the independent couplings is only 5. As a matter 
of fact they can be written in terms of the independent couplings 
$A_1\,,A_2\,,A_4\,,B\,,K$ defined by the formulae 
\ba 
\label{eq:se} &&\frac{3 i}{16\pi^4}\int\,d^4\ell\,\frac{
\ell^\alpha\ell^\beta}{(\ell^2-m^2)[(\ell+q)^2-m^2](v\cdot\ell+\Delta)
(v^\prime\cdot\ell+\Delta)} = A_1g^{\alpha\beta}+\cr&&\cr&&~~~~~~~
~~~~~~~~~~~~~~~~~~~~~~~~~~~~~~~~~+A_2\left( v^\alpha
v^\beta+v^{\prime\alpha}v^{\prime\beta}\right)+A_4\left(
v^{\prime\alpha}v^\beta+v^\alpha v^{\prime\beta}\right),\cr &&\cr&&\frac{3
i}{16\pi^4}\int\,d^4\ell\,\frac{\ell^\alpha}{(\ell^2-m^2)
[(\ell+q)^2-m^2](v\cdot\ell+\Delta)(v^\prime\cdot\ell+\Delta)}
=B\,\left( v^\alpha +v^{\prime\beta}\right),\cr &&\frac{3
i}{16\pi^4}\int\,d^4\ell\,\frac{1}{(\ell^2-m^2)[(\ell+q)^2-m^2]
(v\cdot\ell+\Delta)(v^\prime\cdot\ell+\Delta)}=K \, . 
\ea 
The dependence of  the $g_k$ on these couplings is in Table 2.
\begin{center}
\begin{tabular}{|c|c|}
\hline   Coupling &
CQM formula \\
\hline
 $g_{0}$&$ \beta\,[(4m+2\Delta_H)B-m^2K]$\\
  $g_{1}$&$ \beta\,[2 A_1+2 A_2+2\omega A_4-4mB+m^2K]$\\
 $g_{2}$&$\beta\,[2 A_1+2 \omega A_2+2 A_4-2(1+\omega)mB+m^2K]$\cr
 $g_{3}$&$2 \beta\,[ A_2-A_4-mB]$\\
 $g_{4}$&$\beta\,[2 A_1+2  A_2+2\omega A_4-(1+\omega)mB+m^2K]$\cr
 $g_{5}$&$\beta\,[ 2A_2-mB]$\cr
 $g_{6}$&$\beta\,[2A_1+2A_4-mB+m^2K]$\cr
  $g_{7}$&$\beta mB$\cr
  $g_{7}^\prime$&$g_1+2g_7$\cr
  $g_{7}^{\prime\prime}$&$g_1+g_7$\cr
 $g_{8}$&$2\beta[(1-\omega)A_2-A_4]
 $\cr $g_{9}$&$2\,\beta\,A_4$\cr
  $g_{9}^\prime$&$g_5+g_7$\cr\hline
 \end{tabular}\end{center}\vskip.1cm{\footnotesize  Table 2. Relations
 between  the coupling
constants $g_k$ and the constants $A_1$, $A_2$, $A_4$, $B$ and $K$. Units are
GeV$^{-2}$; $\omega $ is defined in eq.(\ref{14}), with $p^2_1=p^2_2=m^2_D$ and
$p^2=2$ GeV$^2$ for $g_0\,,g_1\,,g_4\,,g_7$ and $p^2=5$ GeV$^2$ for
$g_2\,,g_3\,,g_5\,,g_8\,,g_9$.}\vskip.3cm \noindent In this table \be 
\beta=\frac{m_J^2-p^2}{f_\pi f_J
m_J}\,Z_H\,, \ee and explicit formulae for $A_1\,,A_2\,,A_4\,,B\,,K$ can be
found in Section~\ref{sec.5}. These results have been obtained by a VMD ansatz
similar to Fig.~\ref{fig.1}, but now the l.h.s is modified by the insertion
of a soft pion on the light quark line (with a coupling $q_\pi^\mu/f_\pi
\gamma_\mu\gamma_5$). Let us discuss in some detail one of these couplings,
$g_0$.
 The numerical results for on-shell $D$ mesons, in the soft pion 
limit as a function of the  $J/\psi$ virtuality show a behavior 
similar to that of Fig.~\ref{fig.2}. By the same arguments
 used to determine $g_{JDD}$ in Fig.~\ref{fig.2} we choose  
$p^2=2$ GeV$^2$ and we get 
\be
\frac{g_0}{m_D}=125\pm
15~{\rm GeV}^{-3}~~~~~({\rm soft~pion~limit})\ . 
\label{gddjp} 
\ee 
In order to include hard  pion effects we write the general formula 
\be 
g_k(|\vec q_\pi|)=g_k(0)\,f_k(|\vec q_\pi|),
\ee 
where $f_k(|\vec q_\pi|)$ is a form
factor. We will discuss it in the next section. For the time being we report
the values of all the coupling constants in the soft pion limit in Table 3.
\begin{center}
\begin{tabular}{|c|c||c|c|}
\hline   Coupling &
Results~(GeV$^{-2}$)& Coupling &
Results~(GeV$^{-2}$) \\
\hline
 $g_{0}(0)$&$ +234\pm 45$&   $g_{5}(0)$&$ -274 \pm 49$  \\
  $g_{1}(0)$&$- 235\pm 38$&    $g_{6}(0)$&$ +104\pm 16$  \\
 $g_{2}(0)$&$ -126\pm 19$ &    $g_{7}(0)$&$ +30\pm 5$  \\
 $g_{3}(0)$&$ -252\pm 38$ &     $g_{8}(0)$&$ -106\pm 22$ \\
 $g_{4}(0)$&$ -165\pm 25$ &    $g_{9}(0)$&$ -63 \pm 37 $  \cr
 \hline
\end{tabular}\vskip.1cm
{\footnotesize Table 3. Results for the coupling constants $g_j(|\vec q_\pi|)$
in the soft pion limit $q_\pi\to 0$.}\end{center}\vskip.3cm

In computing  this table we have adopted a criterion of stability in $p^2$
analogous to the one used for $g_0$. For $g_0\,,g_1\,,g_4\,,g_7$
we find
stability at the $J/\psi$ virtuality $p^2=2$ GeV$^2$. For 
$g_2\,,g_3\,,g_5\,,g_8\,,g_9$ we find stability at $p^2=5$ GeV$^2$. 
The technical reason for this difference is that, in the latter case,
the equations do not determine the five constants for  $p^2\approx 0$;
therefore the stability region lies around the center of the $p^2$ 
interval $(0,m_{J}^2)$. For $g_6$ we do not find stability and we derive it by
consistency equations derived from Table 2. To the error arising
from the stability analysis we have added  a further
theoretical uncertainty of $\pm 15\%$ in quadrature. 

An attempt 
to compute quadrilinear couplings {\it via} $SU_4$ symmetry relations 
can be found in~\cite{cina1} and in~\cite{cina2}. For example,
the result for $g_0$ obtained in~\cite{cina2} is 30~GeV$^{-2}$. 
The difference with Table 3 is due to the large $SU_4$ violations 
of our model $(m_c>>m_u,m_s)$.

\section{Form factors}
\label{sec.5}
The coupling constants $g_j$ can be expressed in terms of 
the constants $A_j$,
$B$ and $K$ using the results of Table 2. These constants, for $|\vec
q_\pi|\neq 0$, are expressed in terms of parametric integrals $I_j$ as follows:
\ba 
A_1&=& \frac{C_1(1-\omega^2)-2C_3+2C_2\omega}{2(1-\omega^2)}\ , \cr
A_2&=&\frac{C_1(\omega^2-1)-6C_2\omega+2C_3(2+\omega^2)}{2(\omega^2-1)^2}\ ,
\cr A_4&=&
\frac{C_2(2+4\omega^2)+\omega(C_1(1-\omega^2)-6C_3)}{2(\omega^2-1)^2}\ , \cr
B(q_\pi)&=&\frac{1}{1+\omega}
\int_0^1 dx [I_4(\Delta^\prime(x,q_\pi))-\Delta(x,q_\pi) K(x,q_\pi)]\ ,\\
K(q_\pi)&=&\int_0^1 dx K(x,q_\pi), \ea where \ba C_1(q_\pi)&=& \int_0^1 dx
[I_5(\Delta(x,q_\pi),\Delta^\prime(x,q_\pi), {\omega}) + m^2 K(x,q_\pi)] \ ,\cr
C_2(q_\pi)&=& -I_2-\int_0^1 dx [\Delta^\prime(x,q_\pi)
I_4(\Delta^\prime(x,q_\pi))+\Delta(x,q_\pi) I_4(\Delta(x,q_\pi))\cr&-&
\Delta(x,q_\pi)\Delta^\prime(x,q_\pi) K(x,q_\pi)]\ , \cr C_3(q_\pi)&=&
-\omega I_2 - \int_0^1 dx [(\Delta(x,q_\pi) + {\omega}
\Delta^\prime(x,q_\pi)) I_4(\Delta^\prime(x,q_\pi))
\Delta^2(x,q_\pi)
K(x,q_\pi)]\ , \cr K(x,q_\pi)&=& \frac{\partial}{\partial
m^2}I_5(\Delta(x,q_\pi), \Delta^\prime(x,q_\pi),{\omega}), 
\ea 
and we have defined 
\ba
\Delta (x,q_\pi) &=& \Delta-q_\pi x\ ,\\
\Delta^\prime (x,q_\pi) &=& \Delta-q_\pi x{\omega} \ .
\ea 
To compute the integrals we have applied the Feynman trick with the
shift $\ell+q x \to \ell^\prime$ where the pion momentum is
$q_\mu=(q_\pi,0,0,q_\pi)$.  $\omega$ is computed by eq. (\ref{14})
with the appropriate value of $p^2$, according to the discussion
above. Moreover  we have used the approximation: \be
v^{\prime\mu}= \omega \,v^\mu  \ ,\label{vvo}\ee which has the
correct normalization at $\omega=1$ and also satisfies the
constraint $\omega=v\cdot v^\prime$. Using (\ref{vvo}) we assume
that at least one of the two charmed resonances is off-shell, 
simplifying considerably the numerical computation. 
A numerical calculation of the
integrals in these equations leads to a general fit of the form
factors as follows: \be f_k(|\vec q_\pi|)=\frac{1}{\displaystyle
1+\frac{|\vec q_\pi|}{m_k}}\ , \label{f}\ee with approximately the
same value for all the form factors:\be m_k=0.20\pm0.05~~{\rm
GeV}\ .\label{mk}\ee
\begin{figure}[ht!]
\begin{center}
\epsfig{
        figure=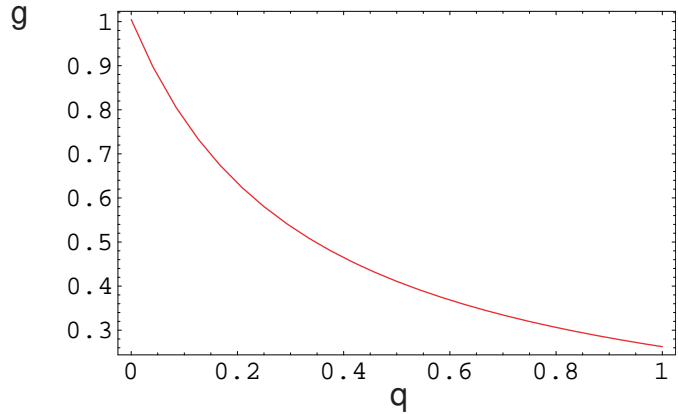}
\caption{\label{fig} \footnotesize \label{fig.g} The dependence of $g(q)\equiv
g_{D^*D\pi}(q)/g_{DD\pi}(0)$ on the pion momentum $q$.}
\end{center}
\end{figure}
It is useful to compute the corrections to the soft pion limit also for the
$DD^*\pi$ coupling constants whose value for $q_\pi\to 0$ was computed in
\cite{nc} and reported in eq. (\ref{ddp}). Using the same technique employed
above we make the substitution \be \frac 1 {(\ell^2-m^2)^2}\to\frac 1{
(\ell^2-m^2)[(\ell+q)^2-m^2]}\ee and compute the integrals appearing in eq.
(63) of \cite{nc} by the Feynman method as above. The result of this
calculation is  plotted  in Fig.~\ref{fig.g}. The dependence can be 
fitted by a formula similar to eq. (\ref{f}) with a mass $m_\chi=0.37$ GeV: 
\be
g_{D^*D\pi}(|\vec q_\pi|)=\frac{g_{D^*D\pi}^{spl}}{1+\frac{|\vec
q_\pi|}{m_\chi}}\ .
\label{37}
\ee

It is useful to compare this result with that found in Ref.~\cite{Isola:2001ar}
for the same quantity computed in a QCD inspired potential model. In that case
one finds again a similar form factor with $m_\chi=0.20\pm 0.10$ GeV.  These
two determinations are sufficiently compatible and induce us to be confident
that the method we have used to get the results (\ref{f}) and (\ref{mk}) is
reliable. It can be noted that considering a different extrapolation, i.e. in  
the pion virtuality $Q^2$, as in Ref. \cite{navarra2}, a smoother behavior
is obtained, which should be relevant in the calculation of the
$J/\Psi$ absorption cross section.

\section{Conclusions}
As discussed  in \cite{falk}, but see also \cite{report}, the leading
contributions to the current matrix element $\langle H(v^\prime)\pi |\bar c
\gamma^\mu c|H(v)\rangle $ in the soft pion limit (spl) are the pole diagrams.
The technical reason is that, in the spl, the reducing action of a pion
derivative in the matrix element is compensated in the polar diagrams by the
effect of the denominator that vanishes in the combined limit $q_\pi\to
0,\,m_c\to\infty$. Let us now compare this result with the effective $JDD\pi$
coupling obtained by a polar diagram with an intermediate $D^*$ state. We get
in this case 
\be
g_0^{polar}\approx\frac{g_{JDD^*}g_{D^*D\pi}m_D}
{2q_\pi\cdot p_D}\ .
\ee 
This expression dominates over the result (\ref{gddjp}) for pion momenta
smaller than 100 MeV. If one restricts the model to the soft pion limit ($|\vec
q_\pi|<100$ MeV), in spite of the rather large value of  $g_0(0)$ the diagrams
containing this coupling are  suppressed, and one can expect a similar result
also for the other couplings. However, to allow the production of a
$D^{(*)}D^{(*)}$ pair by processes (\ref{1}), one has to go beyond the spl,
since the threshold for the charmed meson pair is $|\vec q_\pi|=700-1000~$ MeV.
Our results show that in the CQM model this is indeed possible, by including
computed form factors as in (\ref{f}) and (\ref{37}). Similar form factors were
considered in \cite{cina1}, with a different motivation. Here we have shown
that the CQM model not only allows their computation, but also gives the
general expression for the trilinear and quartic couplings of $J/\psi$ to
charmed mesons and pions. In spite of its model dependent character this seems
to us an interesting result. Applications to the problem of the $J/\psi$
absorption in a nuclear medium will be considered elsewhere.

\section*{Acknowledgements} 

One of us, GN, wishes to thank the Theory Division of
CERN for the kind hospitality. ADP is supported by a M. Curie 
fellowship, contract HPMF-CT-2001-01178.

\end{document}